\documentclass[conference]{IEEEtran}
\IEEEoverridecommandlockouts
\usepackage{cite}
\usepackage{amsmath,amssymb,amsfonts}
\usepackage{algorithmic}
\usepackage{graphicx}
\usepackage{textcomp}
\usepackage{xcolor}
\usepackage{dblfloatfix}    % To enable figures at the bottom of page

\usepackage{fancyhdr}
\fancypagestyle{firstpage}{% Page style for first page
\fancyhf{}% Clear header/footer
\fancyhead[C]{To appear at the 25\textsuperscript{th} Euromicro Conference on
Digital System Design 2022 - DSD '22.} % Header
\fancyfoot[C]{\copyright{}2022 IEEE \hfill}% Footer
}

\begin{document}

\title{Designing Approximate Arithmetic Circuits with Combined Error Constraints}

\author{
     \IEEEauthorblockN{Milan \v{C}e\v{s}ka, Ji\v{r}\'{i} Maty\'{a}\v{s}, Vojtech Mrazek\IEEEauthorrefmark{1}, and Tom\'{a}\v{s} Vojnar}
     \IEEEauthorblockA{ Brno University of Technology, Brno, Czech Republic}
     \IEEEauthorblockA{mrazek@fit.vutbr.cz}
}

\maketitle

\thispagestyle{firstpage}  % arXiv title

\begin{abstract}
Approximate circuits trading the power consumption for the quality of results play a key role in the development of energy-aware systems.
Designing complex approximate circuits is, however, a very difficult and computationally demanding process. 
When deploying approximate circuits, various error metrics (e.g., mean average error, worst-case error, error rate), as well as other constraints (e.g., correct multiplication by~0), have to be considered.
The state-of-the-art approximation methods typically focus on a single metric which significantly limits the  
applicability of the resulting circuits. In this paper, we experimentally investigate how various error metrics and their combinations affect the reduction of the power consumption that can be achieved.
To this end, we extend evolutionary-driven techniques that allow us to 
effectively explore the design space of the approximate circuits. We identify principal limitations when complex error constraints are required as well as important correlations among the error metrics enabling the construction of circuits providing the best-known trade-offs between the power reduction and combined error constraints.
\end{abstract}


\section{Introduction}

\emph{Approximate circuits} are digital circuits that trade functional
correctness (precision of computation) for other design objectives such
as chip area or power consumption~\cite{Mittal:2016}. Such circuits play an important role in 
the development of resource-efficient HW and SW systems, including prominent applications such as image and video processing~\cite{Gupta:2013} or architectures for neural networks~\cite{Mahdiani:2010,mrazek:iccad16}. 
Designing approximate systems is, however, a very complex and time-demanding process trying to find optimal trade-offs between the approximation error and resource savings. Automated methods allowing one to develop such circuits are thus in high demand. A recent overview of techniques for the design and application of approximate circuits is available in~\cite{Reda:Shafique:book:2019}.

There exists a vast body of literature (see, e.g.,~\cite{vasicek:sekanina:tec,abacus:2019,mrazek:date:17,Grater:2016})
demonstrating that evolutionary-based algorithms are able  to automatically design innovative
implementations of complex approximate circuits.
In particular, \emph{Cartesian Genetic Programming} (CGP)~\cite{miller:cgp}
was able to provide high-quality trade-offs among the functional
correctness and the different design objectives~\cite{SekaninaCh:2019}.
The key idea behind these approaches is to iteratively modify the initial (accurate) circuit to achieve lower complexity while keeping the candidate circuit roughly correct.
The correctness is typically expressed using an error metric such as \emph{the mean absolute error} (MAE) or \emph{the worst-case error} (WCE), describing the precision of the approximate circuit.
These error metrics were widely used in the approximate circuit design, but there are a lot of different error metrics, including the error rate (ER), the mean relative error, or the average error close to 0, that are beneficial for certain applications~\cite{jiang:jproc:2020}. 
% https://ieeexplore.ieee.org/stamp/stamp.jsp?tp=&arnumber=9165786
% citace: Various design metrics and analytical approaches are useful for the evaluation of approximate arithmetic circuits [49], [66]–[73].
% 
Besides these primary error metrics, additional constraints on the approximation error have been considered in different application domains, e.g., precise multiplication by 0 is crucial for the approximate inference in neural networks~\cite{mrazek:iccad16}.

\begin{figure}[ht]
    \centering
    \includegraphics[width=\columnwidth]{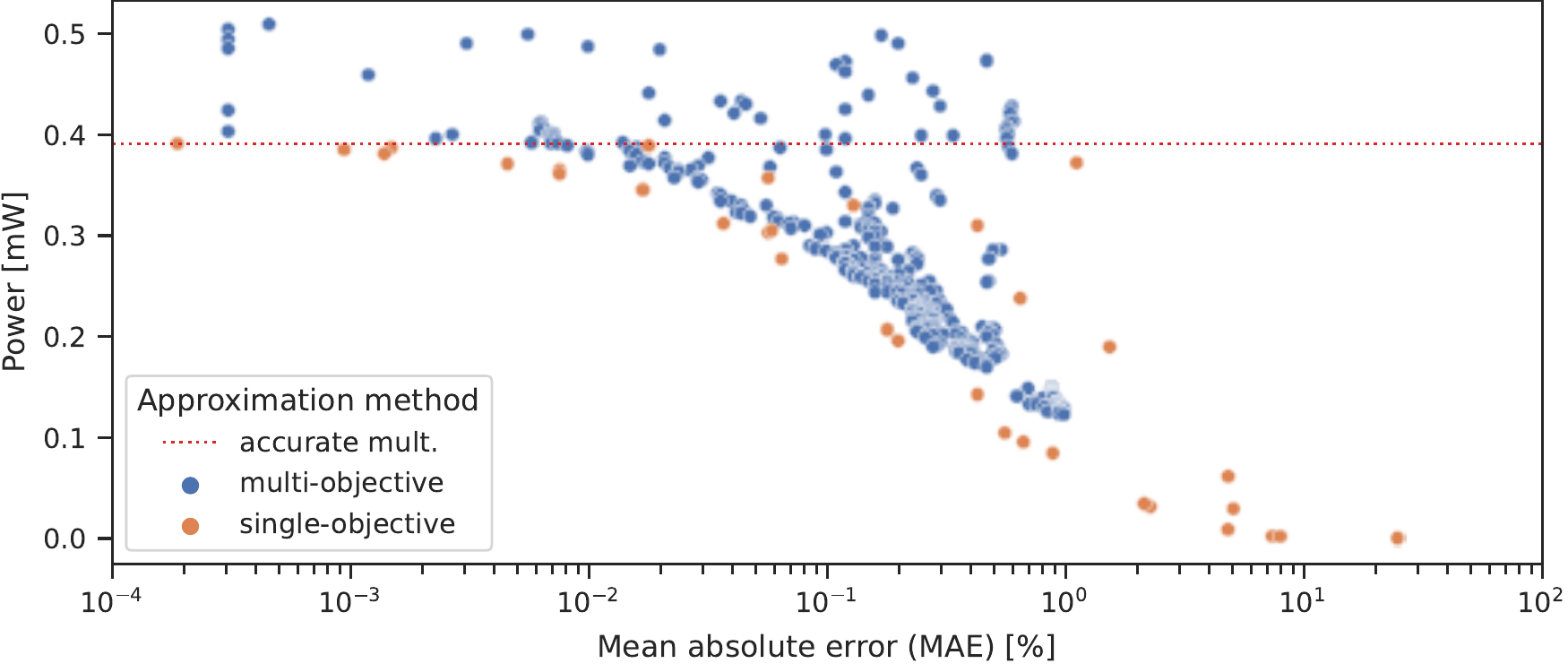}
    \caption{Comparison of final results for single-objective (CGP) and multi-objective (NSGA-II) approximation methodologies}
    \label{fig:so_mo}
\end{figure}

\begin{figure}[ht]
    \centering
    \includegraphics[width=\columnwidth]{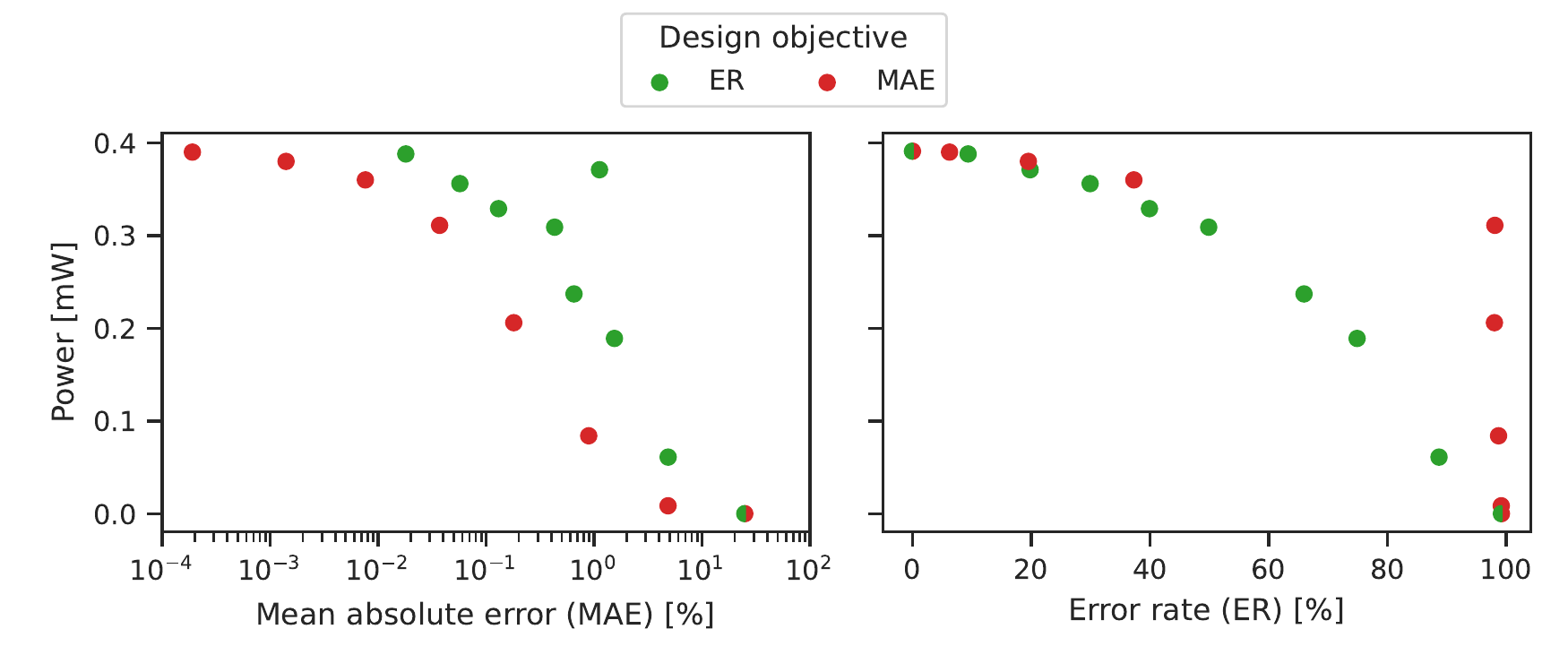}
    \vspace{-2em}
    \caption{The quality of the approximate circuits designed using different error metrics.}
    
    \label{fig:so_pos}
\end{figure}

To support effective deployment of approximate circuits in a broad class of resource-aware applications, various libraries of approximate components, such as \mbox{EvoApproxLib}~\cite{mrazek:date:17}, have been developed. For a given metric and a required error threshold, one can easily select the circuit providing the best power reduction. However, if the target application requires a combination of the error constraints (over different metrics), the existing libraries offer only very limited support. This is given namely by the fact that integrating multi-objective reasoning (such as NSGA-II) into the circuit approximation significantly reduces the performance~\cite{dtis:2016}. As a result,  the single-objective approach achieves considerably better trade-offs -- Fig.~\ref{fig:so_mo} demonstrates this fact in the approximation of 8-bit multipliers. Although there are some results on more complicated error constraints (e.g., in~\cite{mrazek:iccad16}, the authors show that the MAE and the WCE can be effectively combined with the constraint on precise multiplication by 0), there is a gap between state-of-the-art approximation methods and the ability to ensure that multiple error constraints are met. Fig.~\ref{fig:so_pos} demonstrates that single-metric constraints are not sufficient for finding circuits providing the high-quality trade-offs across multiple error metrics. Moreover, it shows that different metrics (and related constraints) have a significantly different impact. The ER constraints lead to circuits where the trade-offs between power reduction and the MAE is not far from the reference trade-offs achieved using the MAE as the design objective (Fig.~\ref{fig:so_pos} left). However, when the circuits are approximated using the MAE, they are not able to achieve useful trade-offs with respect to ER (Fig.~\ref{fig:so_pos} right).

In this paper, we systematically investigate the impact of a broad class of error constraints and their combinations on the possible reduction of the power consumption in approximate circuits. To this end, we extend the single-objective paradigm within the CGP-based approximation allowing us to combine the error metrics in a single fitness function. 
Our aim is to understand which (combination of) constraints permit a practical power reduction and which are too restrictive. Our key observations can be summarized as follows.
\begin{itemize}
    \item There is no single error metric that would ensure high-quality trade-offs between the power reduction and the other metrics.
    \item The ER is antagonistic with respect to the other metrics: using ER as the design objective fails to achieve good results for the other metrics, and using other metrics as the design objective fails to achieve good results for ER.
    \item Combining the ER with the WCE or MAE achieves very good results (typically close to the best-achieved trade-offs) across all monitored metrics.    
\end{itemize}

The approximate circuits constructed using the proposed approach significantly improve the known trade-offs between the accuracy (formulated using various error constraints) and the power reduction. In particular, we obtain circuits outperforming the circuits from EvoApproxLib~\cite{mrazek:date:17}. Although we primarily consider power consumption as a non-functional metric, we expect that similar observations would be obtained for the energy, delay, or the chip area as well.
 
\section{Error metrics}

Accurately assessing the error of an approximate solution is one of the key steps of approximate logic synthesis. Moreover, different target applications of approximate circuits require different approximate behavior. Therefore, various metrics describing the error of approximate circuits have been proposed and shown suitable for different application domains. 

Let $\mathbf{B}=\{0,1\}$. Consider a correct circuit~$G$, also denoted as the \emph{golden circuit}, which computes a~function $f_G$,
and its approximation~$C$, computing a~function $f_C$, where $f_G,f_C:\mathbf{B}^n
\rightarrow \mathbf{B}^m$. 

The primary error metrics for arithmetic circuits are \emph{the mean absolute error} (MAE) and \emph{the worst-case error} (WCE). The WCE is essential when guarantees on the worst behavior of the approximate circuits are required. On the other hand, the mean error better describes the overall error behavior of the approximate solution. 

The WCE is typically defined as follows: 
\begin{equation}
\mbox{WCE}(C,G) = \max_{x\in
\mathbf{B}^n}\left|\mathrm{int}(f_G(x)) - \mathrm{int}(f_C(x))\right| \label{eq:wce}
\end{equation} 
where $\mathrm{int}(x)$ denotes the integer representation of a bit vector~$x$ and
$|i|$ denotes the absolute value of an integer $i$. 

The MAE is an average-based metric similar to WCE. Unlike WCE, MAE averages the values of arithmetic error across all input combinations. It therefore captures the overall error behavior of the approximate circuit.
\begin{equation}
    \mbox{MAE}(C,G) = 2^{-n} \cdot \sum_{x\in \mathbf{B}^n}\left|\mathrm{int}(f_G(x)) - \mathrm{int}(f_C(x))\right| \label{eq:mae}
\end{equation}
To compare different circuits, these metrics are typically relativized to the function output range $2^m$. The relative values are measured in percent.

Besides the two primary metrics, we consider additional metrics and error constraints that are relevant for various applications of approximate arithmetic circuits.

 \emph{The error rate} (ER, also known as error probability) is an error metric that measures the ratio of input vectors for which the output of the approximate circuit is inaccurate (the approximate output differs from the correct output). It is computed as follows:

\begin{equation}
   \mbox{ER}(C,G) = 2^{-n} \cdot \left| \left\{x\in \mathbf{B}^n: f_C(x) \neq f_G(x) \right\} \right| \label{eq:er}
\end{equation}

 \emph{The mean relative error} (MRE, also known as the mean relative error distance) is an error that assumes smaller errors for lower outputs and higher errors for large outputs. To avoid division by zero (if the accurate function output $f_G(x)=0$), value $1$ is considered.
\begin{equation} 
\mbox{MRE}(G,C) = 2^{-n} \sum_{x \in \mathbf{B}^n} \frac{\left|int(f_{G}(x)) - int(f_{C}(x))\right|}{max(int(f_{G}(x)), 1)} \label{eq:mre}
\end{equation}

Another property, important especially for approximate multipliers, is the ability to produce correct results when one of the input operands is 0. Approximate multipliers with this quality have been shown to provide superior performance in applications such as neural networks~\cite{mrazek:iccad16} where the majority of approximate operations is multiplication by 0.

\begin{equation} 
\mbox{ACC0}(G,C) = 
\begin{cases} 

\mbox{1} & \mbox{if } {\forall x \in \mathbf{B}^{n}}: f_G(x) = 0 \Rightarrow  f_C(x) = 0 \\
\mbox{0}  & \mbox{ otherwise.} 
\end{cases}\label{eq:acc0}
\end{equation}

In addition to the above-mentioned metrics measuring the error magnitude, we can also examine other properties of the approximate solution's error distribution. One of them is the mean of the error distribution. The mean tells us whether the approximate solution tends to under or overapproximate the correct result. When the error distribution mean is close to 0, the error introduced over multiple approximate computation steps has the potential to average out. For example, truncated multipliers always underapproximate the correct result, and therefore their error distribution is heavily biased to the negative side. 

\begin{equation} 
    \mbox{AVG}(C,G) = \frac{\sum_{x\in \mathbf{B}^n} (\mathrm{int}(f_G(x)) - \mathrm{int}(f_C(x)))}{2^n} \label{eq:avg}
\end{equation}
Note that, in contrast to the MAE, there is no absolute value in the definition of AVG.

It was shown that approximate multipliers with a Gaussian distribution of the errors are beneficial in many applications as they can be easily emulated by adding a Gaussian noise~\cite{marchisio:date20}. Therefore, we consider a more complex constraint ensuring that the distribution 
of the errors is below a Gaussian distribution $D_{\sigma}$ given by the standard deviation $\sigma$ and the mean~0. Note that error 0 is indeed excluded from this condition. This formulation allows for an iterative approximation and forces the introduced errors to be distributed according to $D_{\sigma}$ (i.e., large errors can be introduced only for a small set of inputs).

\begin{equation} 
    \mbox{Gauss}_{\sigma}(G,C) = 
\begin{cases} 
\mbox{1} & \mbox{if } \mathrm{distr}(f_G(x) - f_C(x)) \leq_{\sigma} D_{\sigma} \\
\mbox{0}  & \mbox{ otherwise.}
\end{cases}\label{eq:gauss}
\end{equation}
where $\mathrm{distr}(f_G(x) - f_C(x))$ denotes the distribution (i.e., histogram) of the errors over all $x\in \mathbf{B}^n$ and $\leq_{\sigma}$ is evaluated on intervals of the size $\sigma$ where the number of errors are averaged.

\section{Automated design of approximate circuits}

 In this paper, we focus on functional approximation where the original circuit is replaced by a less complex one that exhibits some output errors but improves non-functional circuit parameters such as the power consumption or the area on the chip. An automated functional approximation is typically formulated as an iterative design space exploration. It starts with an original (exact) circuit and modifies its structure to obtain several candidate approximate circuits. The candidate circuits are then evaluated in terms of functional (approximation error) 
 and non-functional requirements (electrical parameters).
There exist several heuristic approaches to perform the search including SALSA~\cite{salsa:2012}, AXILOG~\cite{axilog:2015}, ASLAN~\cite{aslan:2014}, or ABACUS~\cite{abacus:2019}. As shown in literature~\cite{ceska:iccad17, SekaninaCh:2019}, more advanced search techniques such as various forms of evolutionary algorithms provide significantly better~results.

\subsection{Cartesian Genetic Programming}
% Pouzit text z mutacniho paperu
Since its introduction, Cartesian Genetic Programming (CGP) remains one of the most powerful evolutionary techniques in the domain of logic synthesis and optimization~\cite{miller:cgp,Miller:2019}.
Its adoption in the area of approximate computing seems to be natural 
because circuit approximation can be formulated as an
optimization problem where the error and non-functional circuit parameters are
conflicting design objectives~\cite{vasicek:sekanina:tec,mrazek:date:17}.

In the logic synthesis and optimization domain, a linear form of CGP is preferred today.
In this case, CGP models a candidate circuit having $n_i$ primary inputs and $n_o$ primary outputs as a linear 1D array of $n_n$ configurable nodes.
Each node has $n_a$ inputs and represents a single gate with up to $n_a$ inputs.
Two-input and single-output nodes are typically used. The inputs can be
connected either to the output of a node placed in the previous $L$ columns, for
some given $L$, or directly to primary inputs. This avoids feedback connections. The function of a node can be chosen from a set $\Gamma$ consisting of $|\Gamma|=n_f$ functions.
Depending on the function of a node, some of its inputs may become redundant.
In addition to that, some of the nodes may become redundant because they are not
referenced by any node connected to a primary output. The fixed number of nodes
$n_n$ does not mean that all the nodes are effectively used. 
This property is a direct consequence of the redundant encoding used in CGP, which enables the existence of neutral mutations. 

\begin{figure}[ht]%
\centering
\includegraphics[width=1\columnwidth]{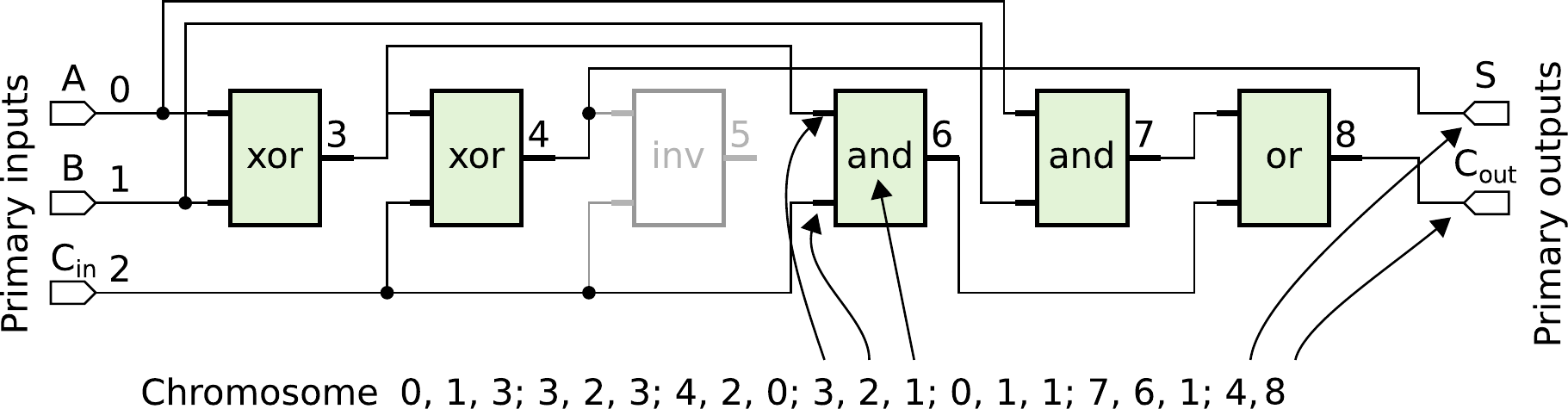}
\caption{An example of a one-bit full adder encoded using the 1D CGP with parameters: $n_i=3$, $n_o=2$, $n_n=6$, $n_a=2$, $\Gamma=\{\mathrm{inv~(encoded~with~0), and~(1), or~(2), xor~(3)}\}$. 
The adder is encoded as five active nodes. One node (node 5) is inactive.}
\label{fig:chromosome}
\end{figure}

The candidate circuits are encoded as follows.
Each primary input as well as each node is associated with a unique index. 
Each node is encoded using $n_a+1$ integers $(x_1,\cdots,x_{n_a},f)$ where the
first $n_a$ integers denote the indices of its fan-ins, and the last integer determines the function of that node.
Every candidate circuit is encoded using $n_n (n_a+1) + n_o$ integers where the last $n_o$ integers specify the indices corresponding with each primary output.
Fig.~\ref{fig:chromosome} illustrates the CGP encoding on an example full adder circuit.

\subsection{Error-oriented design}

Various search strategies have been developed in the context of the evolutionary-driven approximation~\cite{SekaninaCh:2019}.
The majority of the currently available methods are error-oriented in the sense that all logic optimizations leading to an approximate solution are constrained by a predefined {\em error criterion}.
It means that the error determining the quality of the candidate approximate circuits is used as a design constraint. Only non-functional circuit parameters such as power consumption, delay, or area on the chip are subject to optimization and are considered in the cost function.
Depending on the application, the error can be expressed by various metrics, as discussed above.

The majority of the work in the literature formulates the design of approximate
circuits as a single-objective optimization problem despite its
multi-objective nature~\cite{Reda:Shafique:book:2019}. The multi-objective CGP and its ability to identify non-dominated solutions is indeed a very promising tool to reason about multiple error metrics~\cite{vasicek:sekanina:tec}. In practice, however, it is computationally less expensive to execute a single objective CGP optimizing a given parameter several times and having the remaining ones as the constraints~\cite{SekaninaCh:2019}.

Given a candidate circuit $C$, the fitness function $f(C)$ of the single-objective error-oriented method is defined as follows: 
\begin{equation}
%\begin{align*} 
f(C) = \begin{cases} \mathrm{cost}(C) & \mbox{ if
} \mathrm{error}(G, C) \leq \mathcal{T}, \\ \infty  & \mbox{ otherwise.} \end{cases}
%\end{align*}
\label{eq:fitness}
\end{equation}
Here, $\mathrm{cost}(C)$ denotes the cost of the candidate approximate circuit~$C$
in terms of electrical properties that are of interest in a particular context
(for example, in our experiments, we use the power consumption) and 
$\mathrm{error}(C)$ is the output of an error metric that reflects the quality
of the approximation with respect to the original accurate implementation $G$.

\section{Proposed experimental methodology}
\begin{figure}[ht]
    \centering
    \includegraphics[width=0.8\columnwidth]{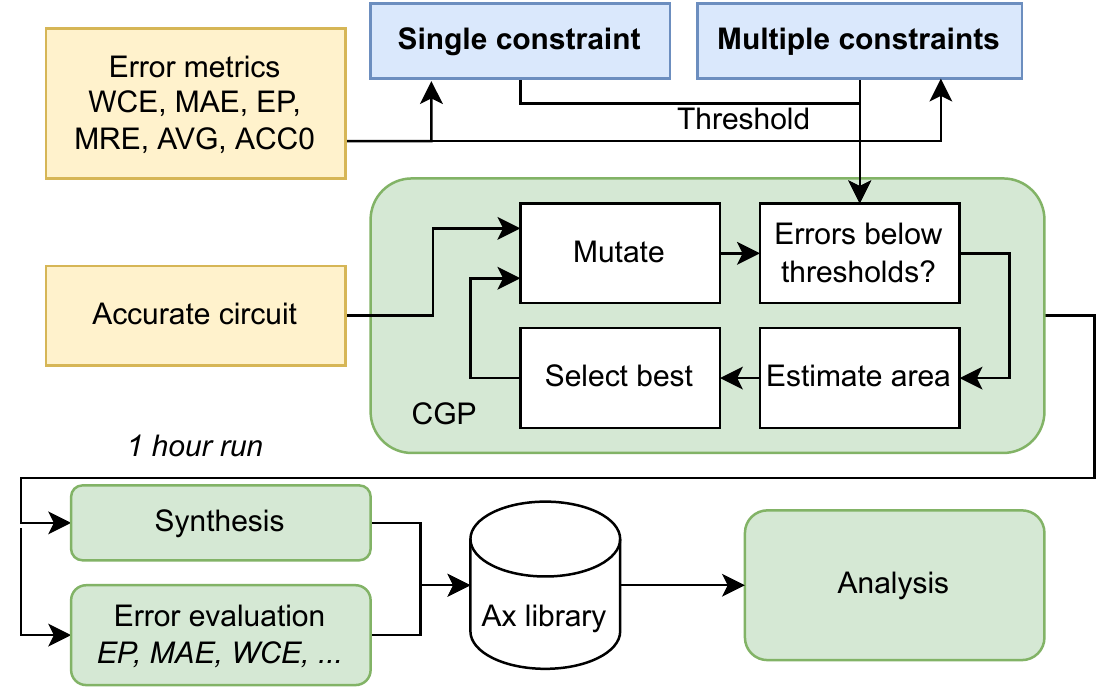}
    \caption{Experimental setup for automated approximation evaluation}
    \label{fig:experiments}
    \vspace{-1em}
\end{figure}

In order to systematically investigate the impact of particular error constraints and their combinations on the power reduction of approximate circuits, we extend the state-of-the-art CGP-based approximation techniques. 
First, we consider additional error metrics including MRE, ER, AVG and Gauss. Second, we extend the fitness function to reason about multiple constraints within a single fitness function. In particular, we extend the predicate in Equation~\eqref{eq:fitness} as follows:
\begin{equation}
    \bigwedge_{i=0}^n  error_i(G, C) \leq \mathcal{T}_i \label{eq:error}
\end{equation}
The overall experimental setup is illustrated in Fig.~\ref{fig:experiments}.

In our experimental evaluation, we focus on 8-bit multipliers. As the golden circuit, we use the multiplier from the \textit{yosys} Verilog star (*) operator implementation. Note that 8-bit multipliers provide important building blocks for more complicated circuits. Thus, their approximation has been intensively explored using various approaches (i.e., they are at the core of almost all experimental suits for circuit approximation). Our results can be naturally generalized to different circuits and bit widths. For structurally simpler circuits (e.g., adders), a larger power reduction is typically achieved for the same error constraint, but accordingly to our preliminary experiments, the dependencies on the error metrics are similar. Approximation of more complicated circuits suffers from a time-demanding evaluation of the error~\cite{ceska:iccad17}, and thus it is much harder to sufficiently explore the design space. 

For each considered configuration of the error metrics and constraints $\mathcal{T}_i$, 10 independent runs of the CGP were executed. The time-out for each run was set to one hour, enabling a sufficient exploration of the design space of the 8x8 multipliers~\cite{mrazek:iccad16}. The CGP algorithm uses the standard parameters as follows: $n_n=400, n_i=16, n_o=16$. % JM: v CGP chromozomu je 400 uzlu => n_n = 400
To determine the precision of the circuit with respect to the given metrics, we simulated all $2^{16}$ input combinations using 64-bit arithmetic vectorization. The final circuits were synthesized using the FreePDK 45~nm technological library to determine their power consumption. 

The experiments were performed at a computer cluster with 14-core Intel Xeon E5-2680 CPUs. In total, we ran more than 27,000 independent CGP runs that took more than 80 \mbox{CPU-days}.

\section{Experimental Evaluation}
The goal of the experimental evaluation is to answer the following research questions that help us to understand how particular error constraints and their combinations affect the power reduction of the corresponding approximate circuits.

\emph{Q1: Is there a single error metric allowing us to design circuits having good trade-offs across multiple metrics?} Existence of such a metric would indicate an important correlation among the metrics simplifying the design of multi-purpose approximate circuits.    

\emph{Q2: Can we effectively combine different error metrics?} How does the combination affect the power reduction that can be achieved?  

\emph{Q3: Is there a combination of error metrics allowing us to design circuits having good trade-offs across multiple metrics?} Or is the resulting constraint too restrictive to achieve a useful power reduction?

%\subsection*{Q1: Approximability with respect to particular error metrics}

% {
% \color{magenta}
% The main message:

% %The single-metric CGP is able to find high-quality circuits for the following metrics: MAE, WCE (known from literature), and MRE and ER. As expected, the constraints on AVG and ACC0 lead to degenerated circuits implementing trivial functions (e.g., constantly return 0) that are not practically useful.
% }

\subsection*{Q1: Is there a single error metric ensuring global quality?}
\begin{figure}[ht]
    \centering
    \includegraphics[width=\columnwidth]{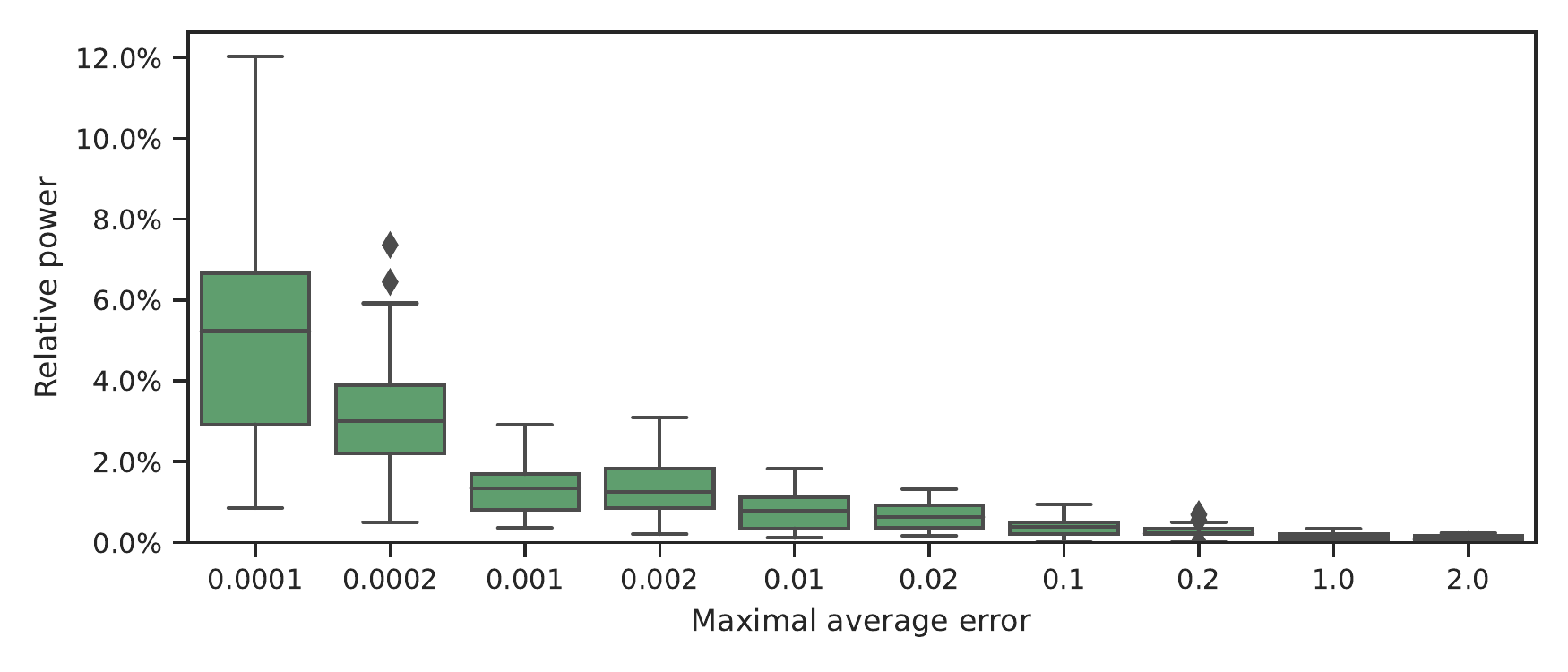}\vspace{-1em}
    \caption{Distribution of relative power if only average error is constrained}
    \label{fig:avg_single}
    \vspace{-0.8em}
\end{figure}

As expected, constraints on the binary metric ACC0 lead to degenerated circuits implementing trivial functions (e.g., constantly return 0). Similarly, constraining only the average error (AVG) produces highly approximated circuits where a significant part of the circuit logic is removed even for very small error thresholds (see Fig.~\ref{fig:avg_single}). These experiments clearly demonstrate that pure ACC0 or AVG constraints do not lead to practically useful circuits.

\begin{figure}[ht]
    \centering
    \includegraphics[width=\columnwidth]{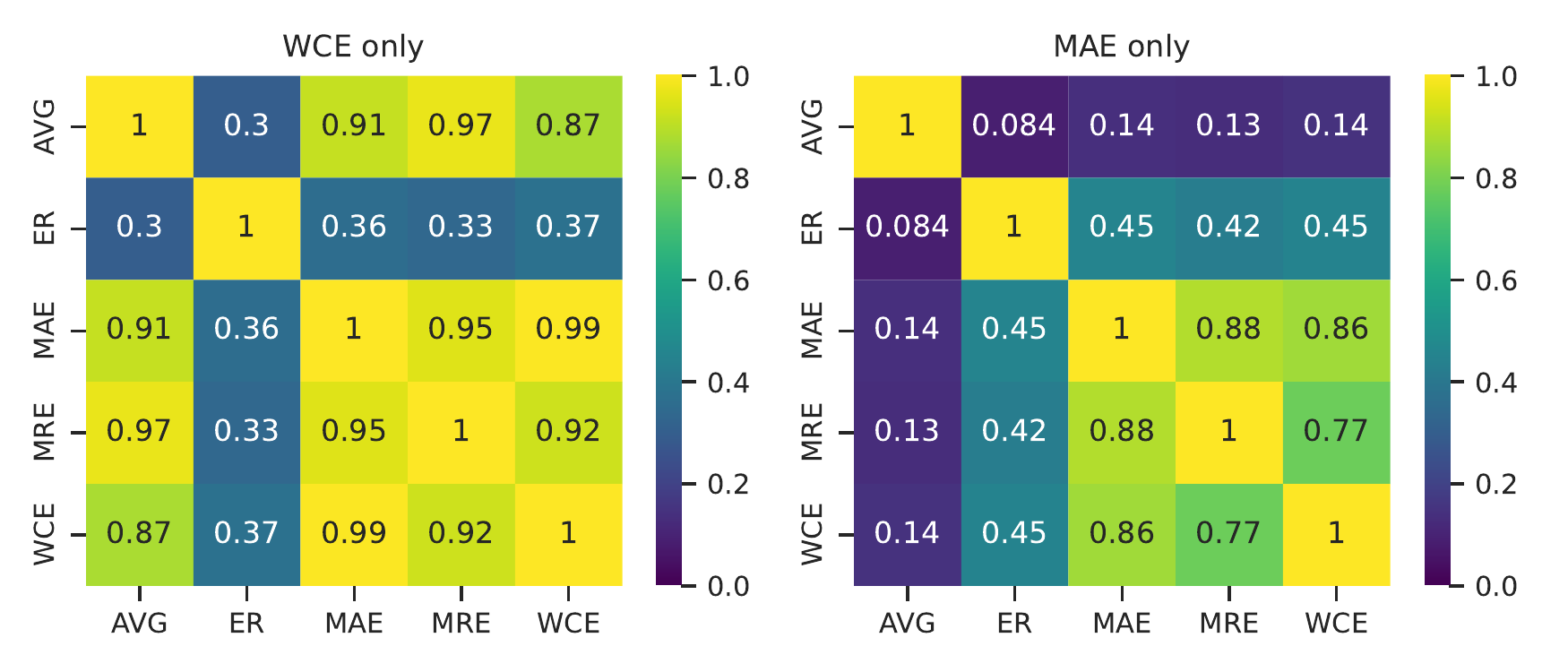}\vspace{-1em}
    \caption{Correlation (absolute value of Pearson number) of the considered error metrics for the circuits obtained using the WCE constraints (left) and the MAE constraints (right).}
    \label{fig:pearsonr}
    
    %\vspace{-0.5em}
\end{figure}

From the existing literature~\cite{mrazek:date:17,ceska:iccad17}, we know that CGP implementing the error-oriented approach is able to find high-quality circuits for the MAE as well the WCE. We first investigate if there exist some correlations among the considered metrics in these circuits. Fig.~\ref{fig:pearsonr} lists absolute values of the Pearson numbers between the particular pairs of metrics. In circuits obtained using the WCE constraints, we observe that all metrics are highly correlated except of the ER. In circuits obtained using the MAE constraints, the correlation is significantly lower and vanishes for the AVG. Our results also indicate a strong correlation between the MAE and WCE, in particular, if we limit the MAE to a value $x$, the WCE never reaches a value larger than 3.2$x$.

\begin{figure}[ht]
    \centering
    \includegraphics[width=\columnwidth]{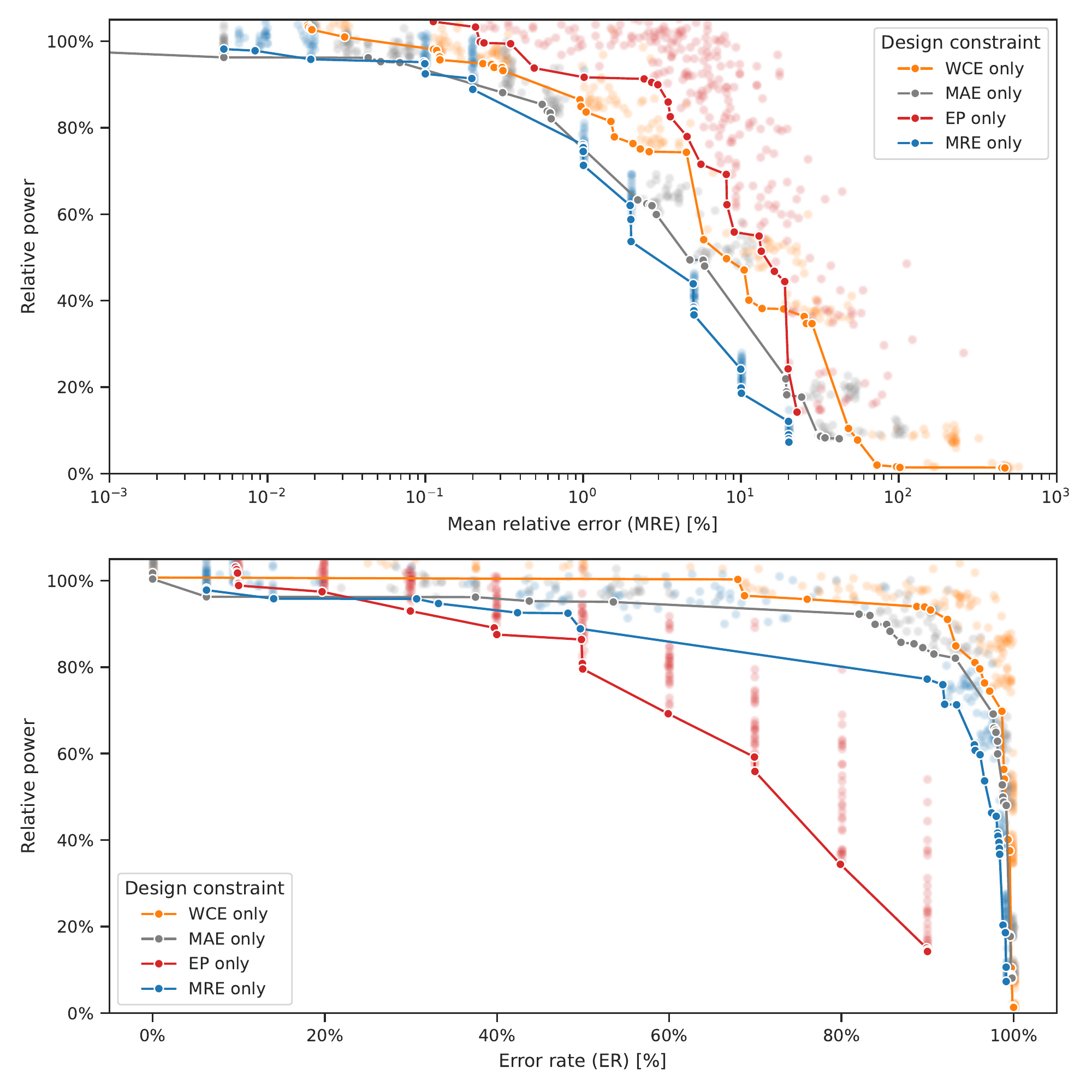}\vspace{-1em}
    \caption{Achieved trade-offs with respect to the MRE (top) and the ER (bottom) when particular design objectives are used.}
    \label{fig:single_obj}
    \vspace{-0.8em}
\end{figure}

Despite of these interesting correlations, we show that none of the considered metrics ensures the global quality of the circuits. It means that optimizing the circuits with respect to any of these metrics does not ensure (close to) optimal trade-offs between the power reduction and (some of) the other metrics.
Fig.~\ref{fig:single_obj} demonstrates this fact by plotting the trade-offs for the MRE and the ER. Observe in the bottom part of the figure (showing the trade-offs with respect to the ER) that the circuits optimized with respect the ER (the red line) are significantly below the other lines. It means that the other metrics do not ensure a good quality with respect to the ER. On the other hand, the circuits optimized with respect the ER  are significantly above the other lines in the upper part of the figure (showing the trade-offs with respect to the MRE). It means that the ER does not ensure a good quality with respect to the MRE. A more systematic view is presented in Fig.~\ref{fig:compare}.

Figures~\ref{fig:single_obj} and \ref{fig:compare} show not only the Pareto-sets (visualized by the lines) but all circuits obtained for the given metric and considered thresholds (visualized by the blurred points). This shows, in contrast to the preliminary study on EvoApproxLib (Fig. \ref{fig:so_pos}), that not only selected but all designed circuits have poor performance with respect to the global quality. In other words, this means that not only the Pareto-optimal solutions do not achieve good performance, but we have not found any circuit with good trade-offs across multiple metrics.

\subsection*{Q2: Can we effectively combine the error metrics}

In this section, we investigate whether adding a secondary error constraint to the fitness function can help CGP to find high-quality circuits with respect to multiple error metrics. In particular, we combine the primary error metrics from the literature (MAE and WCE) with the other metrics defined in Section 2. We do not combine the MAE and the WCE as they are strongly correlated.

\begin{figure}[ht]
    \centering
    \includegraphics[width=\columnwidth]{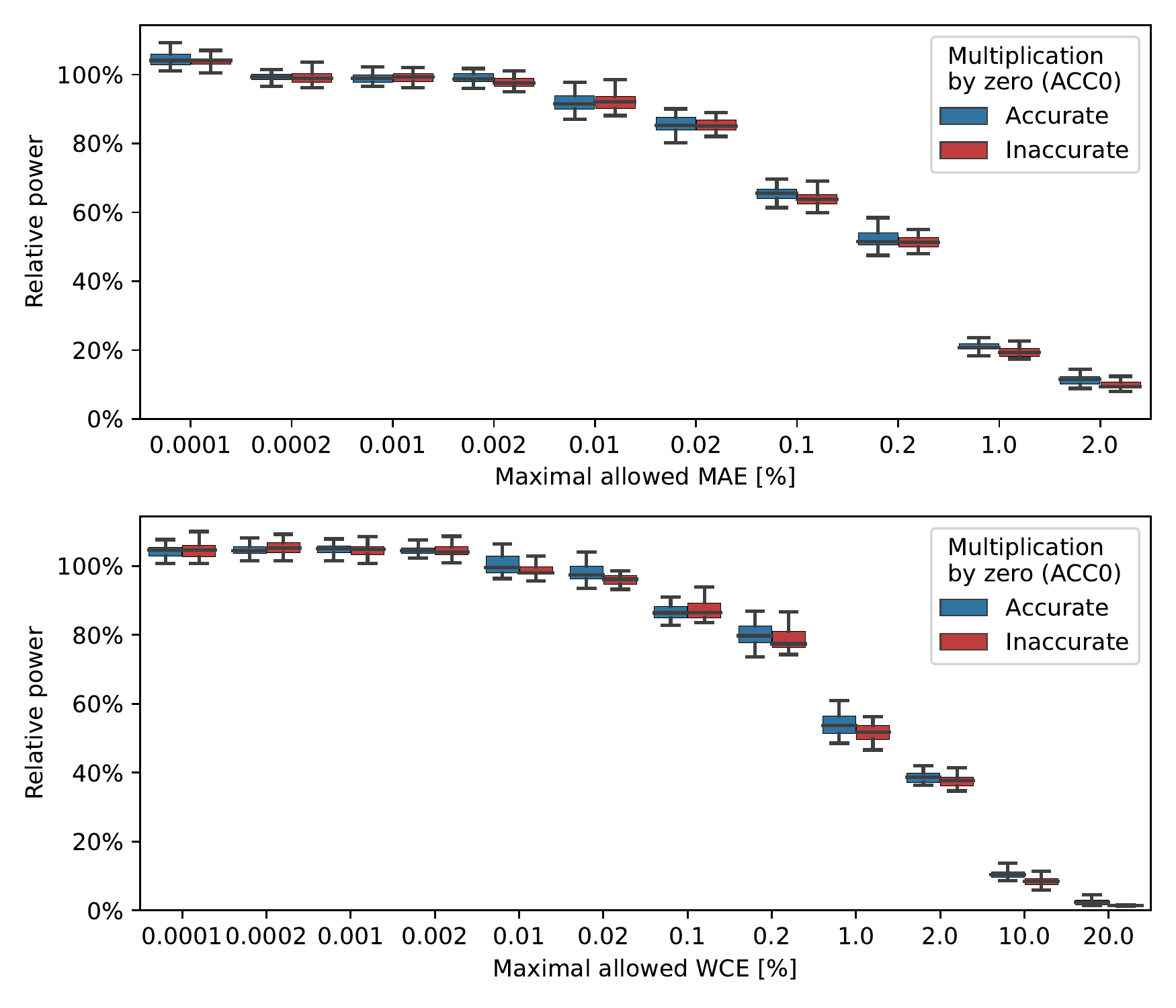}\vspace{-1em}
    \caption{Power consumption of all designed circuits with and without the constraint on accurate multiplication by zero (ACC0).}
    \label{fig:acc0_box}

\end{figure}

In Fig.~\ref{fig:acc0_box}, we first confirm the hypothesis from~\cite{mrazek:iccad16} that adding the ACC0 constraint does not affect the trade-off between the power consumption and the WCE and the MAE, respectively. Adding this constraint changes the median of relative power consumption of designed circuits only about less than 1\% on average. Moreover, in 50\% configurations, this difference is not statistically significant according to the Mann-Whitney U-test.

\begin{figure}[ht]
    \centering
    \includegraphics[width=0.95\columnwidth]{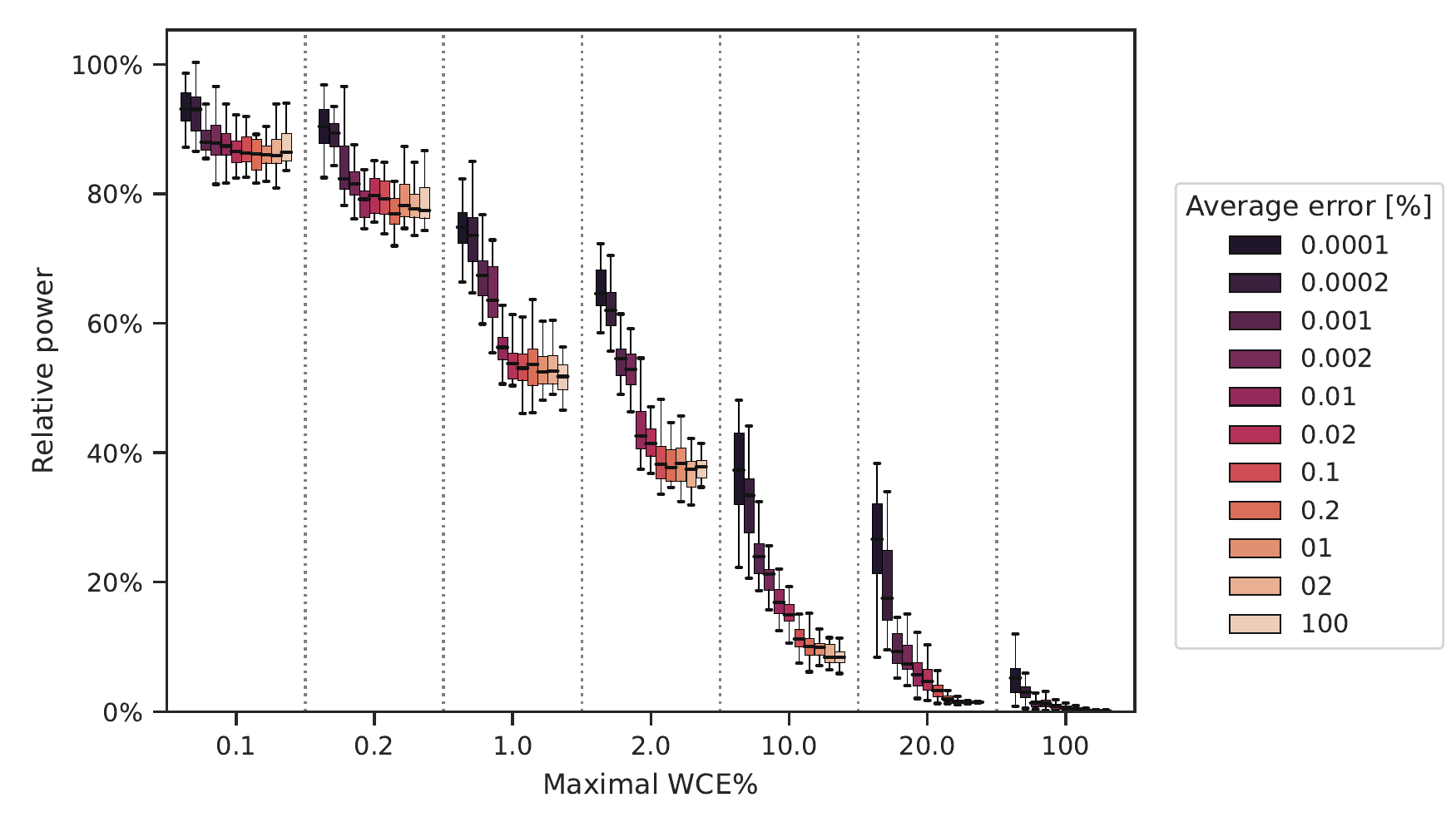}\vspace{-1em}
    \caption{Relative power of circuits designed with limited maximal WCE and AVG errors in the fitness.}
    \label{fig:avg_box}
    \vspace{-0.4em}
\end{figure}

Contrary, adding the AVG constraint has a statistically relevant impact on the trade-offs, as shown in Fig.~\ref{fig:avg_box}. If the WCE is below 0.1\%, the impact of the additional AVG constraint is small, but for larger WCEs, we observe a significant drop in the power reduction. It ranges from  3\% when a moderate threshold on the AVG error is required to 20\% if only a small deviation in AVG is allowed.

\begin{figure}[ht]
    \centering
    \includegraphics[width=0.95\columnwidth]{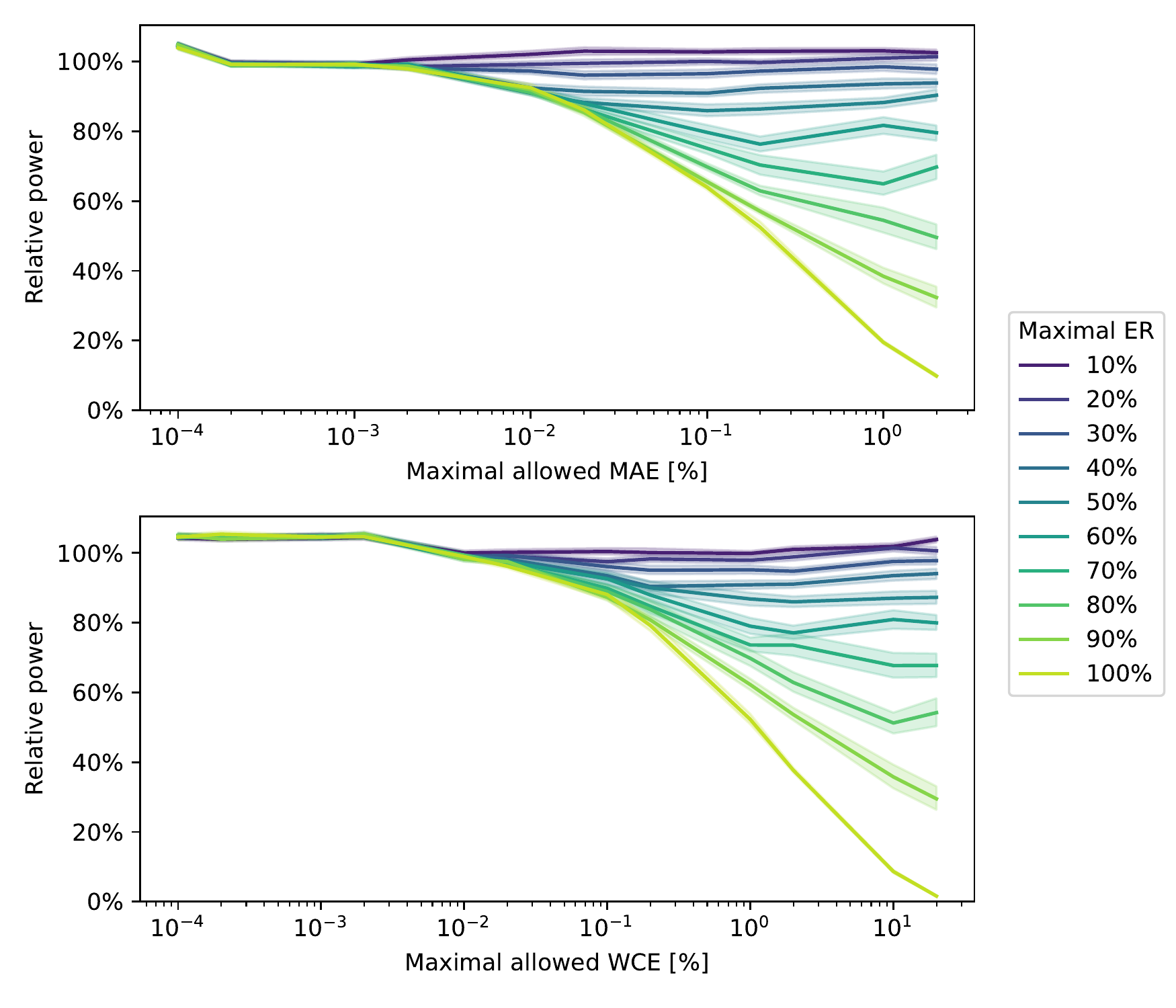}\vspace{-1em}
    \caption{Achieved trade-offs obtained by combining ER with the MAE and the WCE, respectively.}
    \label{fig:er-line}
    \vspace{-1em}
\end{figure}

\begin{figure}[ht]
    \centering
    \includegraphics[width=0.9\columnwidth]{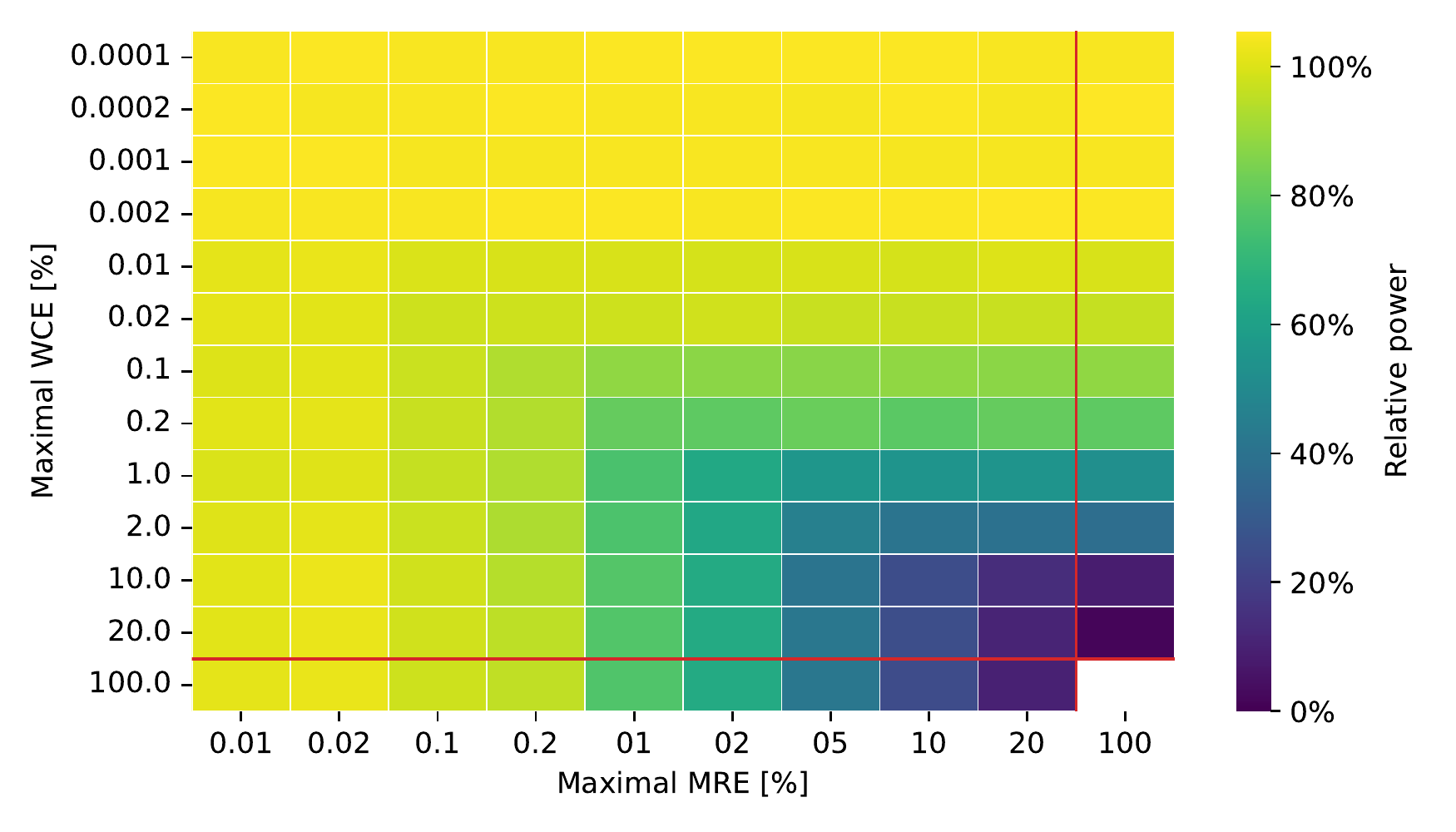}\vspace{-1em}
    \caption{Achieved trade-offs obtained by combining the WCE with the MRE.}
    \label{fig:mre}
    \vspace{-1em}
\end{figure}

An additional constraint on the ER as well as on the MRE has even more significant impact on the achievable power reduction. Fig.~\ref{fig:er-line} shows the results for ER. For example, requiring ER below 70\% prevents achieving power reduction larger than 30\%. This reduction corresponds to the settings where only ER below 50\% is required or only MAE below 0.1\% is required. Surprisingly, requiring  the ER below 70\% prevent achieving high MAE errors (even if the given threshold would permit such an error). For example, for ER=30\%, the maximal MAE was 1.47\%, even if the MAE was not limited. Similar trends are observed when constraints on the MRE are added. Fig.~\ref{fig:mre} shows that, in contrast to ER, the MRE affects the resulting power reduction symmetrically to the WCE. The same observations hold for the combination of the MRE with the MAE (not shown here).

\setcounter{figure}{13}
 \begin{figure*}[b]
    \centering
    \vspace{-1em}
    \includegraphics[width=\textwidth]{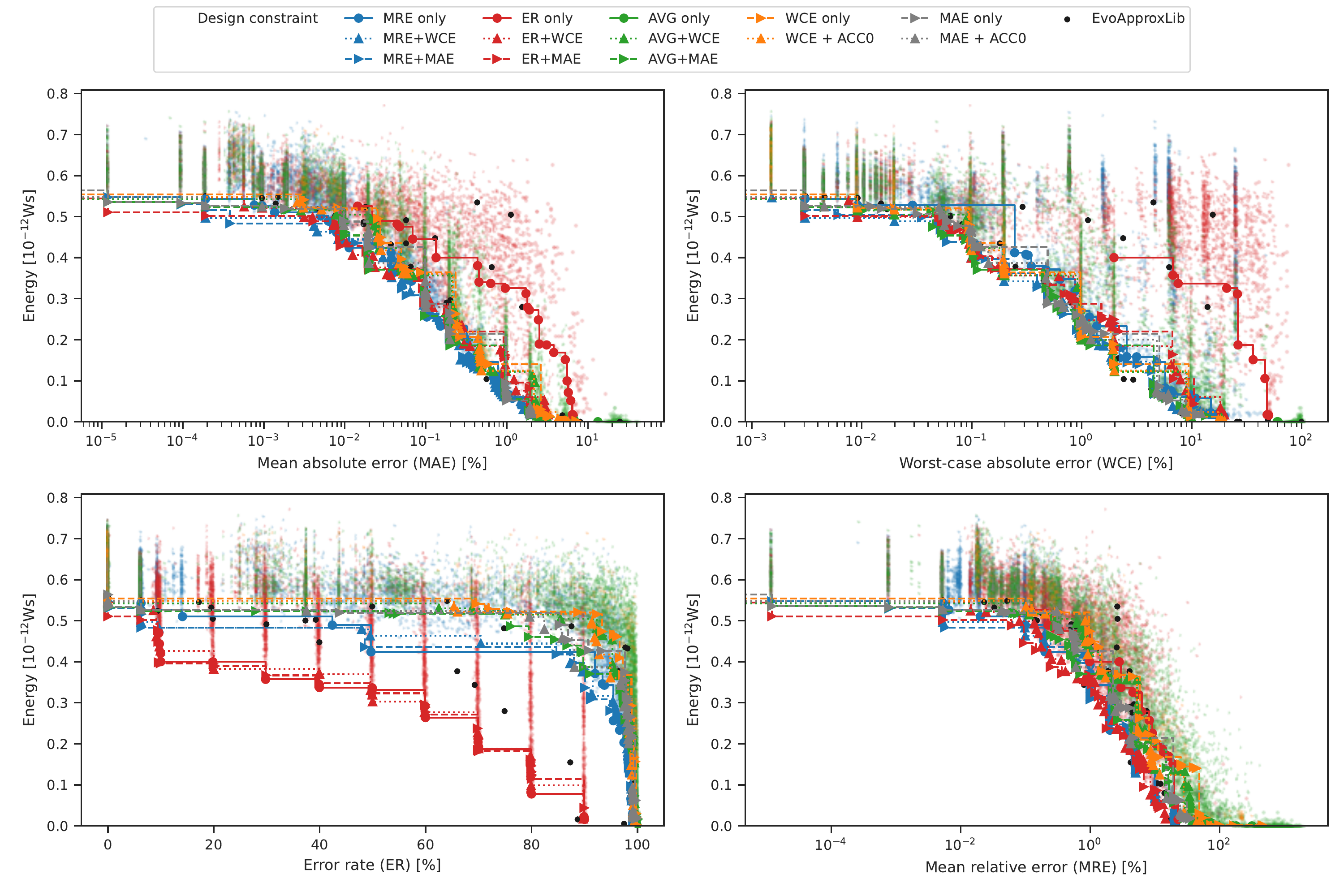}
     \vspace{-2.8em}
    \caption{Achieved trade-offs between the power reduction and the key error metrics -- global comparison of considered deign strategies as well as circuits from \mbox{EvoApproxLib}~\cite{mrazek:date:17}}
    \vspace{-0.8em}
    \label{fig:compare}
\end{figure*}

From these experiments, we can conclude that for some combinations of the metrics, CGP is able to handle combined constraints and find circuits providing useful power reductions. As expected, the combined constraints typically reduce the power reduction compared to the single constraints. 

\setcounter{figure}{11}
\begin{figure}[ht]
     \centering
     \includegraphics[width=0.92\columnwidth]{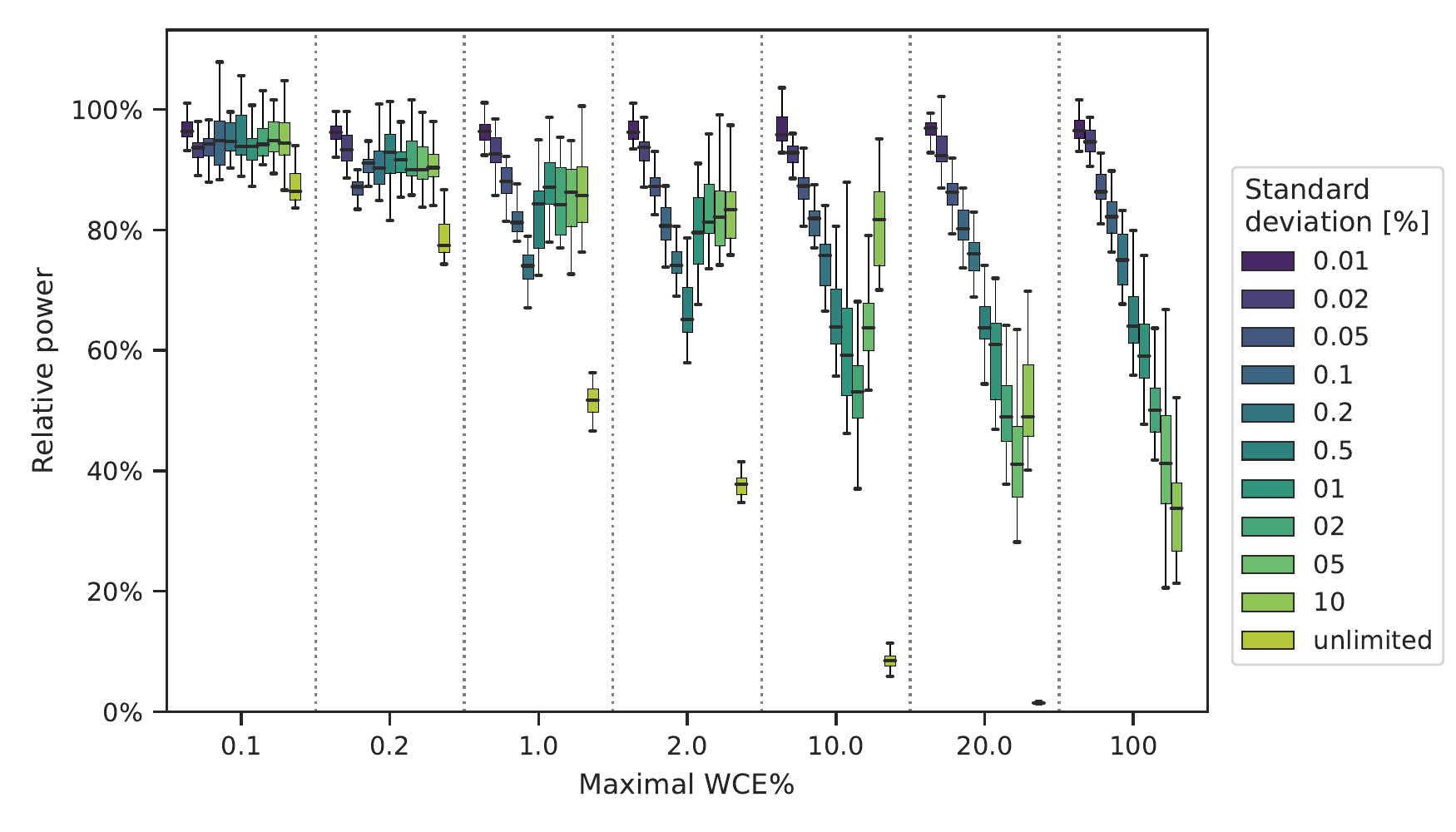}\vspace{-1.5em}
     \caption{Achieved trade-offs obtained by combining the WCE with the constraint on distance from the Gaussian distribution (the Gauss error).}
     \label{fig:gauss_box}
     \vspace{-1.5em}
\end{figure}

The situation is different when a complex constraint on the distribution of the error is required. Fig.~\ref{fig:gauss_box} shows the results for the approximation combining the WCE with the constraint on the error distribution in the rustling circuit. Recall we consider the Gauss constraint (defined in Equation \eqref{eq:gauss}) that is parameterized by the standard deviation $\sigma$. For the majority of the WCE levels, we observe that relaxing the constraint, i.e., increasing $\sigma$ does not necessarily allow for a better reduction (see the V-shapes of the trade-offs). This indicates that CGP struggles to handle this error metric. 
\begin{figure}[ht]
    \centering
    \includegraphics[width=0.92\columnwidth]{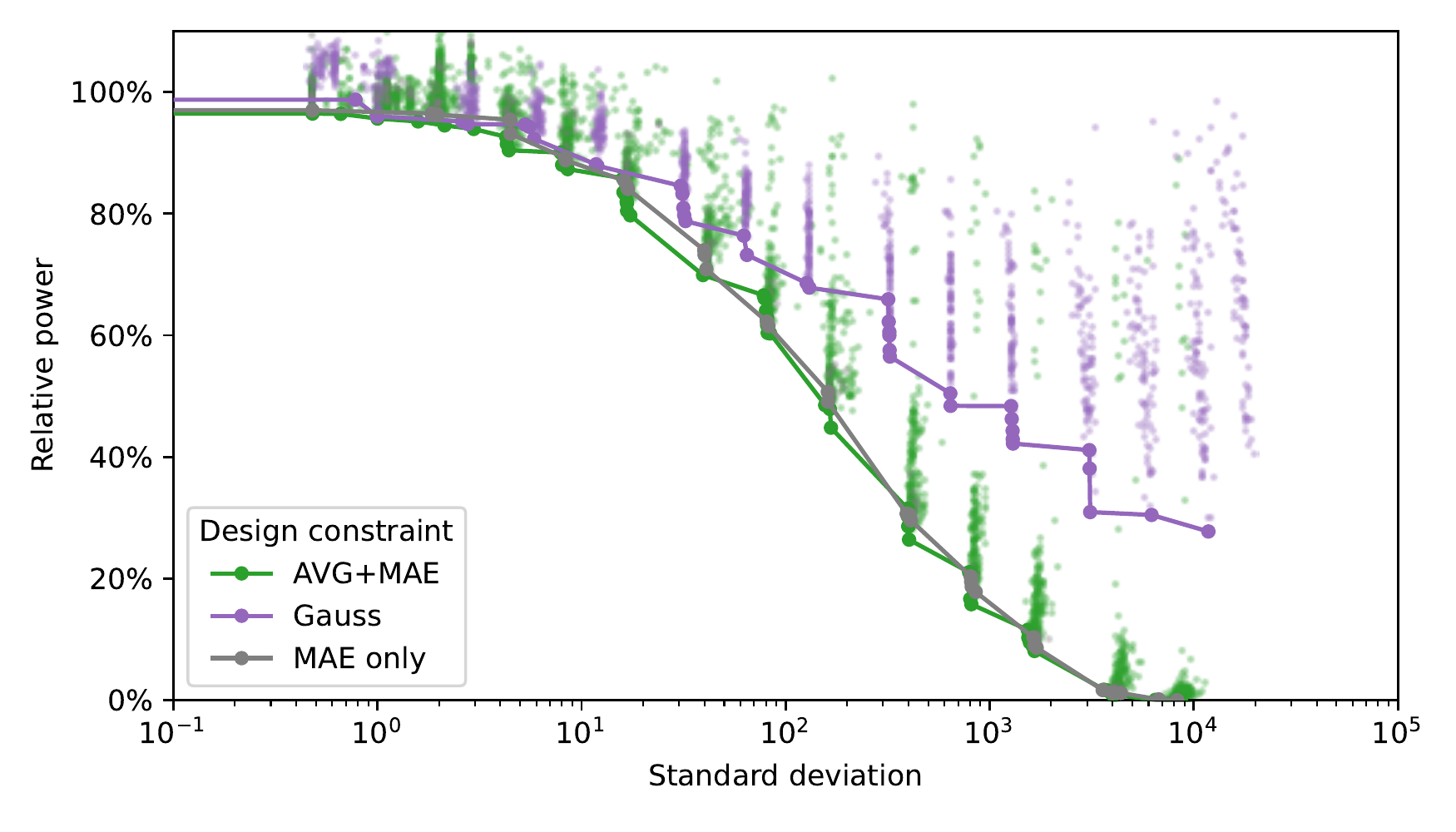}\vspace{-1.5em}
    \caption{Achieved trade-offs between the power reduction and standard deviation of the error for three design constrains.}
    \label{fig:gauss_const}
    \vspace{-1.5em}
\end{figure}

Fig.~\ref{fig:gauss_const} shows the trade-offs between the power reduction and the standard deviation of the errors in circuits obtained using different design objectives (we assume here that the error distribution is close to Gaussian and compare only the standard deviation). We can see that the Gauss constraints lead to significantly lower power reduction. Our experiments further show that if the combination MAE+AVG is used, the resulting circuits have an error distribution that is, in many cases, close to the Gaussian distribution. This indicates that rather than using the complicated Gauss constrain, it is more convenient to use different metrics and then select the solutions with the required error distribution.

\vspace{-0.5em}
\subsection*{Q3: Can a combination of the metrics ensure global quality?}

Finally, we investigate if there is a combination of the error constraints that would enable CGP to find circuits having good trade-offs across multiple error metrics. Fig.~\ref{fig:compare} summarizes the key results of our experimental evaluation. In particular, it shows the Pareto-optimal circuits with respect to their energy consumption and four different error metrics (MAE, WCE, ER, MRE). The particular lines correspond to different (combinations of) error constraints that were used as the designed objective (recall the combined fitness function in Equation \eqref{eq:error}). 

As expected from our previous experiments, achieving good trade-offs for the ER requires using this metric in the design objective. See the left bottom plot where the lines corresponding to the combinations with the ER clearly represent the best (lowest energy) solutions. However, the circuits optimized for the ER only, are very far from the optimal solutions for the MAE and the WCE. In terms of the global quality, the combination ER+WCE and ER+MAE is a clear winner (see the red lines with triangles). The circuits designed using these constraints provide almost optimal trade-offs for the ER and MRE (adding the MAE/WCE constraint to the ER further improves the trade-offs with respect to the MRE). For the MAE and WCE, the circuits slightly lag behind the best circuits obtained using alternative design strategies but still provide very good trade-offs. 
%\mc{A similar trend is observed for the AVG error (not shown here)}. 
If the ER is not relevant for the given application domain, surprisingly, the circuits obtained using the single MRE constraint provide very good traded-offs across the remaining metrics.

We also evaluate the AVG error in the circuits presented in Fig.~\ref{fig:compare}. We observe (not shown here), that thanks to the existing correlations, the combined constrains (including ER+WCE and ER+MAE) lead to circuits where the mean of the error distribution is close to 0.
If the additional constraint on the correct multiplication by~0 is required, it can be added to the MRE  constraint as well as to the combination of ER+WCE and ER+MAE without a significant impact on the resulting energy (we do not present here the resulting Pareto-sets, but recall our previous results on ACC0 presented in Fig.~\ref{fig:acc0_box}). As the combination of ER+WCE and ER+MAE is already very successful, we do not thoroughly investigate more complicated combinations. It would require a large set of approximation runs while there is only a very limited potential to further improve the resulting circuits. 

Fig.~\ref{fig:compare} further demonstrates that using a single MAE or WCE constraint does not provide the best approximation strategy with respect to the global quality. This observation is further backed up by including the best circuits from the \mbox{EvoApproxLib}~\cite{mrazek:date:17} that have been obtained using error-oriented CGP with the MAE or WCE constraints (see the black dots). In all sub-figures, we plot the circuits that are in the Pareto-sets (considering circuits from EvoApproxLib) for one of the four considered metrics. We again see a significant gap when the ER is considered (compare the black dots with red lines in the bottom left sub-figure).

The key conclusion of these experiments is that combining the primary metrics (MAE and WCE) with the ER leads to circuits providing global quality across all the considered metrics and provides significantly better trade-offs compared to circuits obtained using alternative design strategies as well as to the best circuits from the literature.

\section{Conclusion}
We systematically investigate how various error constraints affect the power reduction that can be achieved using CGP-based circuit approximation. The key observation can be summarized as follows: i) The error-oriented CGP is able to handle complex design constraints. However, there are situations where including the constrain explicitly in the design objective is not effective (e.g., when a specific error distribution is required). In such cases, it is better to construct a set of candidate circuits using different design objectives and subsequently select the circuits having the required characterization. ii) Despite existing correlations, there is no single metric allowing CGP to find circuits with good trade-offs across multiple metrics. iii) Combining the MAE or WCE constraint with the ER, however, significantly improves the quality of the resulting circuits across all considered metrics. Moreover, the obtained circuits considerably outperform the existing circuits from the literature in terms of global quality.

\subsection*{Generalization of the presented results.} Although our observations build on a single approximation technique (i.e., error-oriented single-objective CGP optimization) and consider primarily only 8-bit multipliers, there is a strong evidence that they can be generalized to other settings: The alternative approximation techniques such as SALSA~\cite{salsa:2012}, SASIMI~\cite{sasimi}, or BLASYS~\cite{hashemi2018blasys} leverage a similar iterative exploration of the design space and typically lag behind the CGP-based approximation. Approximation of the structurally more complex circuits is computationally more demanding, in particular, due to evaluation of the approximation error (incomplete evaluation or advanced formal verification techniques have to be typically used~\cite{ceska:iccad17}). Therefore, the CGP, as well as alternative approaches, would need significantly more time (compared to the one-hour runs we considered) to sufficiently explore the design space. However, from the existing results for more complex circuits, it can be expected that the correlations among the particular error metrics will have similar trends as we observed for 8-bit multipliers. 

\section*{Acknowledgement} This work has been supported by the Czech Science Foundation grant \mbox{GJ20-02328Y} and the FIT BUT internal project FIT-S-20-6427.

\bibliographystyle{IEEEtran}
\bibliography{dsd22}

\end{document}